\documentclass[prl,aps,twocolumn,showpacs]{revtex4}
\usepackage{dcolumn}
\usepackage{graphicx}
\usepackage{amssymb}
\begin{document}

\title{Effect of Metal Element in Catalytic Growth of Carbon Nanotubes}

\author{Oleg V. Yazyev}
\email[]{oleg.yazyev@epfl.ch}
\affiliation{Ecole Polytechnique F\'ed\'erale de Lausanne (EPFL),
Institute of Theoretical Physics, CH-1015 Lausanne, Switzerland}
\affiliation{Institut Romand de Recherche Num\'erique en Physique
des Mat\'eriaux (IRRMA), CH-1015 Lausanne, Switzerland}
\author{Alfredo Pasquarello}
\affiliation{Ecole Polytechnique F\'ed\'erale de Lausanne (EPFL),
Institute of Theoretical Physics, CH-1015 Lausanne, Switzerland}
\affiliation{Institut Romand de Recherche Num\'erique en Physique
des Mat\'eriaux (IRRMA), CH-1015 Lausanne, Switzerland}

\date{\today}

\begin{abstract}
Using first principles calculations, we model the chemical vapor 
deposition (CVD) growth of carbon nanotubes (CNT) on 
nanoparticles of late transition (Ni, Pd, Pt) and coinage (Cu, Ag, Au) 
metals. The process is analyzed in terms of the binding of mono- and 
diatomic carbon species, their diffusion pathways, and the stability 
of the growing CNT.  We find that the diffusion pathways can be controlled 
by the choice of the catalyst and the carbon precursor.
Binding of the CNT through armchair edges is more favorable than 
through zigzag ones, but the relative stability varies significantly 
among the metals. Coinage metals, in particular Cu, are found to favor 
CVD growth of CNTs at low temperatures and with narrow chirality 
distributions.
\end{abstract}

                
\maketitle

Carbon nanotubes (CNTs) are expected to become an 
important constituent of many technologies, in particular of future 
generation electronics \cite{Baughman02}. The inability of performing
growth of CNTs with predetermined chirality indices, and thus electronic
properties, is the major obstacle on the way to incorporating 
CNTs into electronic devices. The chemical vapor deposition (CVD) 
growth of CNTs catalyzed by metallic nanoparticles  
is believed to be the most promising approach
for reaching this goal. This method has already been demonstrated to allow 
for CNT growth at low temperatures \cite{Cantoro06} and to yield limited 
distributions of nanotube chiralities \cite{Bachilo03,Miyauchi04}. 

For a long time, the catalyst composition has been limited to
iron-group metals (Fe, Co, Ni) and their alloys. Recently, it has been 
shown that metallic nanoparticles of many other metals,
heavy late transition (Pd, Pt) and inert coinage metals (Cu, Ag, Au),
are also able to act as catalysts in the CVD growth process \cite{Takagi06,Zhou06}. 
However, these metals show different physical and chemical properties 
\cite{Okamoto00} as a result of the variation of lattice constants, 
$d$-band energies and occupations \cite{Hammer96}, and relativistic 
effects \cite{Hakkinen02}. 
Therefore, important questions arise. What role does the 
chemical composition of the metal catalyst play in the CVD growth process? 
To what extent can the choice of the metal catalyst 
be optimized for achieving deterministic growth of CNTs?

The decomposition of gas-phase carbon precursors on the nanoparticle 
surface is the first step of the CVD growth of CNTs and 
can easily be controlled via the thermochemistry. This initial step 
is followed by two important processes: (i) the diffusion of carbon on the 
nanoparticle surface or across its interior, and (ii) the nucleation 
of a graphitic fragment followed by the further incorporation of 
carbon into the growing nanotube.
The diffusion process (i) has been shown to be a rate-limiting 
step \cite{Hofmann05,Bartsch05}, while the chirality of the growing CNT is 
decided upon the nucleation process (ii) \cite{Reich05,Reich06}.
The catalyst can be either crystalline or 
liquid \cite{Bartsch05} during the growth process but
only solid metallic nanoparticles of definite shape can act 
as templates for the growth of CNTs with specific chiralities \cite{Zhu05}.
Sufficiently low temperatures are thus necessary for maintaining the crystalline 
state of nanoparticles during the nucleation stage. 
Certain technological applications of CNTs also impose restrictions
on the processing temperatures \cite{Cantoro06}.

In this work, we investigate the role of the chemical composition of the catalytic 
nanoparticles in the CVD growth of CNTs using a total-energy scheme.
In our study, we consider late transition (Ni, Pd, Pt) and coinage (Cu, Ag, Au) metals.
The CVD growth process is analyzed by modeling the diffusion and nucleation steps.
For each considered metal, the binding energies of carbon atoms and dimers on the 
facets as well as in the interior of the nanoparticles are compared. 
For these carbon species, we then calculate activation energies considering both surface 
and bulk diffusion pathways. From the stability of zigzag and armchair edges of the 
growing CNT, we infer the preference for definite chiralities. Our results show 
that coinage metals, in particular Cu, offer attractive opportunities for growing 
CNTs at low temperatures and with narrow chirality distributions.

The surface and the interior of metallic nanoparticles were modeled by means of periodic 
two-dimensional slab and three-dimensional bulk models, respectively. 
Infinite surfaces were shown to be good models for facets, even 
in the case of the smallest metallic nanoparticles \cite{Zhang04}. We primarily considered 
the lowest energy (111) facets \cite{Vitos98}, which dominate the surfaces of 
fcc-metal nanoparticles \cite{Baletto05}.
The surfaces were modeled by three-layer slabs with $p$(3$\times$3) unit cells, 
which guarantee convergence of relative binding energies within 0.1~eV. 
We also considered (100) microfaceted surface step-edges, which we modeled 
by a (311) stepped surface of the same thickness.
This model is also a good approximation for extended (100) nanoparticle facets,
corresponding to the second largest fraction of the surface area of fcc-metal 
Wulff clusters.  
In the two-dimensional models, the periodically repeated slabs were separated 
by at least 10~\AA\ to avoid spurious interactions.
The bulk metals were modeled with 3$\times$3$\times$3 supercells. 

The total energy calculations were performed within a selfconsistent 
density functional theory framework, as implemented in the \texttt{PWSCF} 
plane-wave pseudopotential code of the \texttt{Quantum-ESPRESSO} distribution \cite{QE}. 
The Perdew-Burke-Ernzerhof exchange-correlation density functional \cite{Perdew96} was used. 
The ultrasoft pseudopotentials \cite{Vanderbilt90}
employed in the present study treat the $d$-electrons of both
transition and coinage metals as valence electrons. 
The one-electron valence wave functions and the electron density
were described by plane-wave basis sets with kinetic energy cutoffs
of 25~Ry and 300~Ry, respectively \cite{pasquarello92}.
The Fermi-Dirac function ($k_{\rm B} T$=0.01~Ry) was used for the population 
of the one-electron states in  metallic systems. 
For the slab models, a 4$\times$4$\times$1 ${\mathbf k}$ point sampling was performed.
The bulk supercells were sampled with 4$\times$4$\times$4 ${\mathbf k}$ point meshes. 
The atomic positions of the bottom layer were kept fixed in 
ideal bulk positions during the relaxation in the slab models. 
The nudged elastic band method with the climbing-image scheme \cite{Mills95} 
was used to find the transition states of the diffusion paths.

\begin{figure}
\includegraphics[width=8.5cm]{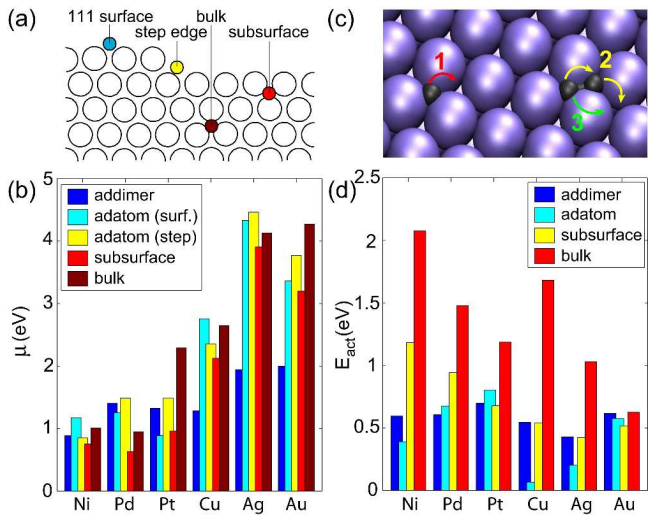}
\caption{\label{fig1} (Color online) (a) Schematic representation of studied carbon positions at 
metal surfaces. (b) Chemical potentials of monoatomic and diatomic carbon in different positions
expressed with respect to the chemical potential of carbon in graphene.
(c) Surface diffusion mechanisms for the carbon adatom (\textbf{1})
and the carbon dimer (\textbf{2} -- concerted mode; \textbf{3} -- atom-by-atom mode) 
on a (111) fcc metal surface and (d) respective diffusion barriers for different metals.
}
\end{figure}

{\it Binding of monoatomic and diatomic carbon.}
The chemical potential of different forms of carbon expressed with respect
to the total-energy per atom of graphene $\mu_{\rm G}$ is defined as   
\begin{eqnarray}
\mu =\frac{E - E_{\rm M}}{n_{\rm C}} - \mu_{\rm G},
\end{eqnarray}
where $E$ and $E_{\rm M}$ are the total energies of the systems with and without 
carbon, respectively, and $n_{\rm C}$ corresponds to the number of carbon atoms.
For reference, the chemical potential of the free carbon atom in the spin-triplet ground 
state is 7.7~eV. 

We considered the binding and the diffusion of both carbon atoms and dimers,
which are initially produced upon the dehydrogenation of common carbon sources 
like CH$_4$, C$_2$H$_4$, C$_2$H$_2$, and C$_2$H$_5$OH.
For all considered positions [Fig.~\ref{fig1}(a)] and metals, the chemical potentials of monoatomic and 
diatomic carbon are positive, i.e.\ indicating lower stability than for graphene, 
and are generally higher for the inert coinage metals [Fig.~\ref{fig1}(b)]. On coinage metals, 
the surface dimers are considerably more stable than all forms of monoatomic carbon.
This implies that the diffusion process only involves surface dimers, when the 
carbon source corresponds to a diatomic carbon precursor such as C$_2$H$_2$.
At variance, atomic carbon has the lowest chemical potential for late transition metals,
as a subsurface interstitial in Ni and Pd and as a surface adatom at the Pt surface.

Adatoms on a (111) surface are threefold coordinated and occupy two different positions,
ccp and hcp~\cite{Zhang04}, with chemical potentials differing by less than 0.1~eV. 
Only the lowest adatom chemical potentials are shown in Fig.~\ref{fig1}(b). 
The highest adatom chemical potential was obtained for Ag, in agreement with 
the calculations of Abild-Pedersen {\it et al.}\ \cite{Abild-Pedersen07} This low stability 
results from the energy position of the $d$-band, which is lowest for this metal \cite{Hammer96}. 
At surface steps, the carbon adatom assumes a semi-octahedral five-fold coordination showing 
the same local neighborhood as at an adsorption site on the (100) surface. 
For Ni and Cu the adsorption site at the step is preferred over that on a (111) terrace,
as found for Ni in a previous study \cite{Bengaard02}. However, the converse is true 
for all heavier metals showing larger lattice constants.

Carbon atoms in subsurface and bulk positions occupy either tetrahedral
or octahedral voids, in which they show coordination numbers of 4 and 6, 
respectively. Subsurface carbon atoms located in voids closest to the surface 
are more stable than surface carbon adatoms for all metals except Pt.
The preferred occupation sites are the tetrahedral voids in Au and Pt 
characterized by large lattice constants, and octahedral voids in the 
lighter metals. Only the lowest-energy values are given in Fig.~\ref{fig1}(b).
Bulk interstitials, i.e.\ carbon atoms located deep inside the crystalline 
nanoparticle, have considerably higher chemical potentials than subsurface 
interstitials due to an enhancement of the elastic response energy
\cite{Todorova02,Abild-Pedersen06}. Bulk interstitials preferentially
reside in octahedral voids for all metals studied.
Subsurface and bulk dimers are found to be highly unstable and
are not further discussed here.

{\it Activation barriers of diffusion.}
The relevant diffusion channels of monoatomic and diatomic carbon [Fig.~\ref{fig1}(c)] on metallic nanoparticles 
are addressed through the calculation of their activation barriers $E_{\rm act}$ [Fig.~\ref{fig1}(d)].
For monoatomic carbon, we considered surface, subsurface and bulk diffusion modes. In case of diatomic
carbon, only the surface diffusion mode was investigated. The reported barrier is 
the total-energy difference between the transition state and the lowest of the connected minima.

The adatom diffusion barrier is lower for the coinage metals with respect to the late-transition 
metal belonging to the same row. Moving down the periodic table results in higher diffusion barriers.
For the adatom diffusion on the Ni(111) surface, we found a barrier of 0.39~eV, in agreement
with previous theoretical estimates \cite{Helveg04,Zhang04,Hofmann05,Abild-Pedersen06}.
This value is also in fair agreement with the activation energy of the experimental CNT growth
process, suggesting that surface diffusion is the rate-limiting step \cite{Hofmann05,Bartsch05}.
The adatom diffusion on Cu and Ag surfaces is characterized by surprisingly low barriers of 
0.07~eV and 0.20~eV, respectively.  This implies that a significant acceleration of the diffusion 
step can be achieved for these metals when a monoatomic carbon precursor (e.g.\ CH$_4$) is used.

Subsurface and bulk diffusion occur via hopping of interstitial carbon atoms between adjacent 
tetrahedral and octahedral voids. The diffusion barriers for this channel is
generally higher than for adatom diffusion. However, the activation energy for subsurface diffusion
decreases with increasing lattice constant, following an opposite trend with respect to the adatom 
diffusion and leading to competition between the two diffusion modes for the heaviest metals.
Activation barriers of subsurface diffusion are always lower than those of the corresponding bulk 
diffusion due to the reduced elastic response in proximity of the nanoparticle surface. 
However, in the particular case of Au, activation energies for surface, subsurface, and bulk 
diffusion fall in a narrow range between 0.52 and 0.64 eV, implying that monoatomic carbon 
in gold can diffuse uniformly across the nanoparticle already at low temperatures.

When the carbon precursor gives rise to diatomic carbon, the dimer diffusion channel becomes relevant.
We considered two different mechanisms for the surface dimer diffusion, the concerted and the atom-by-atom 
modes, as illustrated pictorially in Fig.~\ref{fig1}(c). 
For the metals studied, the concerted mode shows activation barriers exhibiting small variations.
Calculated values range between 0.43~eV for Ag and 0.70~eV for Pt, and are comparable to the 
diffusion barrier of the carbon adatom on Ni(111).  The atom-by-atom mode yields higher 
diffusion barriers with energy differences ranging from almost zero for Pt to 0.62 eV for Ni.
In Fig.\ \ref{fig1}(d), the activation energies reported for dimer diffusion correspond to the 
lowest energy mode, i.e.\ to the concerted mode.

\begin{figure}
\includegraphics[width=8.5cm]{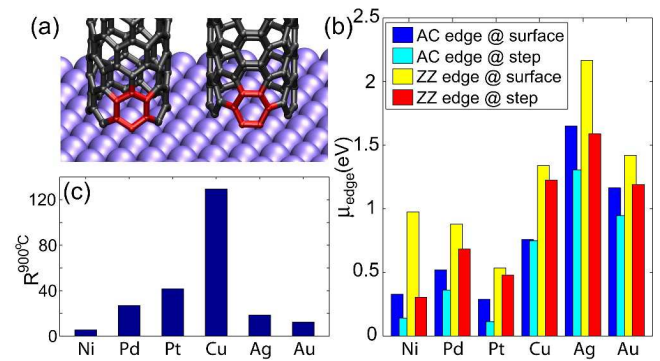}
\caption{\label{fig2} (Color online) (a) Schematic representation of the binding of 
ZZ (left) and AC (right) carbon nanotube to a flat (111)
surface of the catalyst nanoparticle. The edge fragments used 
to model the interaction between the nanotube and the nanoparticle 
surface are shown in red. (b) Chemical potentials of edge atoms of AC 
and ZZ fragments bound to the metal surfaces and steps.
The chemical potential of carbon atoms at free AC and ZZ 
edges are 2.13~eV and 2.83~eV, respectively. (c) Thermodynamic 
chirality preference ratios at a synthesis 
temperature of 900$^\circ$C for various metal catalysts.}
\end{figure}

{\it Binding of CNT to metal nanoparticle.}
%
To understand the relation between the chirality of the growing CNT and
the chemical composition of the catalyst, a comprehensive study of the 
interaction between graphitic fragments and metallic nanoparticles is required.
However, such an investigation is computationally prohibitive
because of the large configuration space involved \cite{Reich05,Reich06}. 
To address this issue, we here focused on the binding of 
minimal structural units of armchair (AC) and zigzag (ZZ) edges to either 
flat or stepped surfaces. 
We modeled AC and ZZ edges by 1,2-didehydrobenzene (C$_6$H$_4$) 
and dehydrobenzene (C$_6$H$_5$) molecules having 2 and 1 edge atoms, 
respectively, as illustrated in Fig.~\ref{fig2}(a).

The binding energies are quantified by the chemical 
potential per edge atom $\mu_{\rm edge}$ defined as 
\begin{eqnarray}
\mu_{\rm edge} =\mu^{\rm free}_{\rm edge} + \frac{E - E_{\rm frag} - E_{\rm M}}{n_{\rm C}},
\end{eqnarray}
where $\mu^{\rm free}_{\rm edge}$ is the chemical potential per edge atom for 
an unbound edge, $E_{\rm frag}$ the total energy of an isolated model fragment, 
and $n_{\rm C}$ the number of edge atoms. 
We estimated $\mu^{\rm free}_{\rm edge}$ through the 
calculations of ideal graphene and suitable graphene nanoribbons.
For $\mu^{\rm free}_{\rm edge}$ of AC and ZZ edges, we found
values of 2.13~eV and 2.83~eV, in good agreement with previously 
reported results \cite{Fan03}.

%
The calculated chemical potentials $\mu_{\rm edge}$ corresponding to
optimally placed fragments bound to either facets or step-edges
are given in Fig.~\ref{fig2}(b) for the various metals.
These results show that all studied metals share common 
qualitative features:
(i) The chemical potententials $\mu_{\rm edge}$ are always lower than 
$\mu^{\rm free}_{\rm edge}$, implying that the binding of the graphitic 
fragment to the metallic nanoparticles is energetically favored. 
Binding to step-edges is systematically preferred over binding to flat 
terraces, in accord with experimental indications \cite{Zhu05,Rodriguez-Manzo07}.
(ii) The chemical potentials $\mu_{\rm edge}$ in Fig.\ \ref{fig2}(b)
and those of mono-atomic and di-atomic species in Fig.\ \ref{fig1}(b) exhibit
similar behavior as the metal is varied, consistent with
the properties of the metal $d$-band \cite{Hammer96}.
However, the former chemical potentials are always lower for corresponding 
metals, indicating that the aggregation of bound C atoms in the form 
of graphitic fragments is energetically favored.
(iii) The chemical potentials $\mu_{\rm edge}$ are always positive, i.e.\ the 
C atoms are less stable than in the reference ideal graphene, suggesting that 
there exists a driving force towards extending the size of graphitic aggregates.
The combination of these three features points to the existence of a negative 
chemical potential gradient for the nucleation and further growth of CNTs from 
a carbon feedstock. This supports the catalytic activity of a wide range 
of metals, including the inert coinage metals.
 
%
The chemical potentials $\mu_{\rm edge}$ in Fig.\ \ref{fig2}(b) also provide insight
into the ability of different metal catalysts of growing CNTs with definite
chiralities. For all considered metals, AC edges are found to be more
stable than ZZ edges. In order to provide a more quantitative measure of
this preference at a growth temperature $T$, we define the ratio
\begin{eqnarray}
R^T =\exp\left[-\frac{\mu_{\rm edge}^{\rm AC} - 
\mu_{\rm edge}^{\rm ZZ}}{kT}\right],
\end{eqnarray}
assuming that the CNT nucleation occurs under thermodynamic control.
For a graphitic CNT nucleus, the chemical potential of the edge carbon atoms is 
higher than that of the other carbon atoms, leading to the growth of CNT with 
minimal edge perimeters. Under this assumption, the ratio $R^T$ characterizes 
the distribution of CNT chiralities indicating the preference of growing 
AC-like or ZZ-like CNTs. Ultimate selectivity is achieved for ZZ 
and AC CNTs for $R^T$=$0$ and $R^T$$\rightarrow$$\infty$, respectively.

%
Figure~\ref{fig2}(c) shows $R^T$ for the different metal catalysts at  
a typical CVD synthesis temperature of 900$^\circ$C. All metals show 
a strong preference for growing AC rather than ZZ nanotubes,
in agreement with experimental reports for iron-group transition 
metals \cite{Bachilo03}. In Fig.\ \ref{fig2}(c), $R^{\rm 900^\circ C}$
ranges between 5 for Ni and 130 for Cu. The case of Cu is particularly
noteworthy for several reasons. First, its high value of $R^T$ implies 
that the CNTs grown on Cu nanoparticles are highly enriched by metallic 
AC species, in accordance with preliminary experimental reports 
\cite{Zhou06}. Second, Cu features the lowest carbon diffusion barriers
suggesting that CVD synthesis could take place at much lower temperatures.  
In these conditions, the chirality preference ratio $R^T$ would be further 
enhanced. Third, for Cu, the chemical potentials in Fig.\ \ref{fig2}(b) 
show that graphitic fragments indifferently bind to flat or stepped 
surfaces, suggesting that step-edge formation on Cu nanoparticles is 
not likely to occur during the nucleation of carbon nanotubes \cite{Zhu05,Rodriguez-Manzo07}. 
Preserving the shape of the nanoparticle is a prerequisite for the 
realization of chirality-specific CNT growth on predetermined
catalytic templates.

In conclusion, our study of the diffusion and nucleation stages of CVD growth 
of CNTs highlights the potential of catalytic nanoparticles of coinage metals. 
For these metals, the stability and the diffusion barriers of diatomic 
carbon allow one to restrict the diffusion pathways to the nanoparticle 
surface by choosing an appropriate gas-phase carbon source.
Furthermore, particularly low diffusion barriers are found for carbon 
adatoms in the case of Cu and Ag, suggesting the possibility of 
realizing growth at low growth temperatures. The Cu catalyst is also 
found to be the most promising candidate for growing carbon nanotubes 
of definite chirality. 

We acknowledge useful discussions with L.~Forr\'o, 
K.~Lee, A.~Magrez, and \v{Z}.~\v{S}ljivan\v{c}anin.  We used 
computational resources at CSCS.


\begin{thebibliography}{31}
\expandafter\ifx\csname natexlab\endcsname\relax\def\natexlab#1{#1}\fi
\expandafter\ifx\csname bibnamefont\endcsname\relax
  \def\bibnamefont#1{#1}\fi
\expandafter\ifx\csname bibfnamefont\endcsname\relax
  \def\bibfnamefont#1{#1}\fi
\expandafter\ifx\csname citenamefont\endcsname\relax
  \def\citenamefont#1{#1}\fi
\expandafter\ifx\csname url\endcsname\relax
  \def\url#1{\texttt{#1}}\fi
\expandafter\ifx\csname urlprefix\endcsname\relax\def\urlprefix{URL }\fi
\providecommand{\bibinfo}[2]{#2}
\providecommand{\eprint}[2][]{\url{#2}}

\bibitem[{\citenamefont{Baughman et~al.}(2002)\citenamefont{Baughman, Zakhidov,
  and de~Heer}}]{Baughman02}
\bibinfo{author}{\bibfnamefont{R.~H.} \bibnamefont{Baughman}},
  \bibinfo{author}{\bibfnamefont{A.~A.} \bibnamefont{Zakhidov}},
  \bibnamefont{and} \bibinfo{author}{\bibfnamefont{W.~A.}
  \bibnamefont{de~Heer}}, \bibinfo{journal}{Science}
  \textbf{\bibinfo{volume}{297}}, \bibinfo{pages}{787} (\bibinfo{year}{2002}).

\bibitem[{\citenamefont{Cantoro et~al.}(2006)\citenamefont{Cantoro, Hofmann,
  Pisana, Scardaci, Parvez, Ducati, Ferrari, Blackburn, Wang, and
  Robertson}}]{Cantoro06}
\bibinfo{author}{\bibfnamefont{M.}~\bibnamefont{Cantoro}}  {\it et al.},
  \bibinfo{journal}{Nano Lett.} \textbf{\bibinfo{volume}{6}},
  \bibinfo{pages}{1107} (\bibinfo{year}{2006}).

\bibitem[{\citenamefont{Bachilo et~al.}(2003)\citenamefont{Bachilo, Balzano,
  Herrera, Pompeo, Resasco, and Weisman}}]{Bachilo03}
\bibinfo{author}{\bibfnamefont{S.~M.} \bibnamefont{Bachilo}}  {\it et al.},
  \bibinfo{journal}{J. Am. Chem. Soc.}
  \textbf{\bibinfo{volume}{125}}, \bibinfo{pages}{11186}
  (\bibinfo{year}{2003}).

\bibitem[{\citenamefont{Miyauchi et~al.}(2004)\citenamefont{Miyauchi, Chiashi,
  Murakami, Hayashida, and Maruyama}}]{Miyauchi04}
\bibinfo{author}{\bibfnamefont{Y.}~\bibnamefont{Miyauchi}}  {\it et al.},
  \bibinfo{journal}{Chem. Phys. Lett.} \textbf{\bibinfo{volume}{387}},
  \bibinfo{pages}{198} (\bibinfo{year}{2004}).

\bibitem[{\citenamefont{Takagi et~al.}(2006)\citenamefont{Takagi, Homma,
  Hibino, Suzuki, and Kobayashi}}]{Takagi06}
\bibinfo{author}{\bibfnamefont{D.}~\bibnamefont{Takagi}}  {\it et al.},
  \bibinfo{journal}{Nano Lett.} \textbf{\bibinfo{volume}{6}},
  \bibinfo{pages}{2642} (\bibinfo{year}{2006}).

\bibitem[{\citenamefont{Zhou et~al.}(2006)\citenamefont{Zhou, Han, Wang, Zhang,
  Jin, Sun, Zhang, Yan, and Li}}]{Zhou06}
\bibinfo{author}{\bibfnamefont{W.}~\bibnamefont{Zhou}}  {\it et al.},
  \bibinfo{journal}{Nano Lett.} \textbf{\bibinfo{volume}{6}}, \bibinfo{pages}{2987}
  (\bibinfo{year}{2006}).


\bibitem[{\citenamefont{Okamoto}(2000)}]{Okamoto00}
\bibinfo{author}{\bibfnamefont{H.}~\bibnamefont{Okamoto}},
  \emph{\bibinfo{title}{Desk Handbook. Phase Diagrams for Binary Alloys}}
  (\bibinfo{publisher}{ASM International}, \bibinfo{address}{Materials Park},
  \bibinfo{year}{2000}).

\bibitem[{\citenamefont{Hammer et~al.}(1996)\citenamefont{Hammer, Morikawa, and
  N{\o}rskov}}]{Hammer96}
\bibinfo{author}{\bibfnamefont{B.}~\bibnamefont{Hammer}},
  \bibinfo{author}{\bibfnamefont{Y.}~\bibnamefont{Morikawa}}, \bibnamefont{and}
  \bibinfo{author}{\bibfnamefont{J.~K.} \bibnamefont{N{\o}rskov}},
  \bibinfo{journal}{Phys. Rev. Lett.} \textbf{\bibinfo{volume}{76}},
  \bibinfo{pages}{2141} (\bibinfo{year}{1996}).

\bibitem[{\citenamefont{H{\"a}kkinen et~al.}(2002)\citenamefont{H{\"a}kkinen,
  Moseler, and Landman}}]{Hakkinen02}
\bibinfo{author}{\bibfnamefont{H.}~\bibnamefont{H{\"a}kkinen}},
  \bibinfo{author}{\bibfnamefont{M.}~\bibnamefont{Moseler}}, \bibnamefont{and}
  \bibinfo{author}{\bibfnamefont{U.}~\bibnamefont{Landman}},
  \bibinfo{journal}{Phys. Rev. Lett.} \textbf{\bibinfo{volume}{89}},
  \bibinfo{pages}{033401} (\bibinfo{year}{2002}).

\bibitem[{\citenamefont{Hofmann et~al.}(2005)\citenamefont{Hofmann, Cs{\'a}nyi,
  Ferrari, Payne, and Robertson}}]{Hofmann05}
\bibinfo{author}{\bibfnamefont{S.}~\bibnamefont{Hofmann}}  {\it et al.},
  \bibinfo{journal}{Phys. Rev. Lett.} \textbf{\bibinfo{volume}{95}},
  \bibinfo{pages}{036101} (\bibinfo{year}{2005}).

\bibitem[{\citenamefont{Bartsch et~al.}(2005)\citenamefont{Bartsch, Biedermann,
  Gemming, and Leonhardt}}]{Bartsch05}
\bibinfo{author}{\bibfnamefont{K.}~\bibnamefont{Bartsch}}  {\it et al.},
  \bibinfo{journal}{J. Appl. Phys.} \textbf{\bibinfo{volume}{97}},
  \bibinfo{pages}{114301} (\bibinfo{year}{2005}).

\bibitem[{\citenamefont{Reich et~al.}(2005)\citenamefont{Reich, Li, and
  Robertson}}]{Reich05}
\bibinfo{author}{\bibfnamefont{S.}~\bibnamefont{Reich}},
  \bibinfo{author}{\bibfnamefont{L.}~\bibnamefont{Li}}, \bibnamefont{and}
  \bibinfo{author}{\bibfnamefont{J.}~\bibnamefont{Robertson}},
  \bibinfo{journal}{Phys. Rev. B} \textbf{\bibinfo{volume}{72}},
  \bibinfo{pages}{165423} (\bibinfo{year}{2005}).

\bibitem[{\citenamefont{Reich et~al.}(2006)\citenamefont{Reich, Li, and
  Robertson}}]{Reich06}
\bibinfo{author}{\bibfnamefont{S.}~\bibnamefont{Reich}},
  \bibinfo{author}{\bibfnamefont{L.}~\bibnamefont{Li}}, \bibnamefont{and}
  \bibinfo{author}{\bibfnamefont{J.}~\bibnamefont{Robertson}},
  \bibinfo{journal}{Chem. Phys. Lett.} \textbf{\bibinfo{volume}{421}},
  \bibinfo{pages}{469} (\bibinfo{year}{2006}).
  
\bibitem[{\citenamefont{Zhu et~al.}(2005)\citenamefont{Zhu, Suenaga, Hashimoto,
  Urita, Hata, and Iijima}}]{Zhu05}
\bibinfo{author}{\bibfnamefont{H.}~\bibnamefont{Zhu}}  {\it et al.},
  \bibinfo{journal}{Small} \textbf{\bibinfo{volume}{1}}, \bibinfo{pages}{1180}
  (\bibinfo{year}{2005}).

\bibitem[{\citenamefont{Helveg et~al.}(2004)\citenamefont{Helveg,
  L\'opez-Cartes, Sehested, Hansen, Clausen, Rostrup-Nielsen, Abild-Pedersen,
  and N{\o}rskov}}]{Helveg04}
\bibinfo{author}{\bibfnamefont{S.}~\bibnamefont{Helveg}} {\it et al.},
  \bibinfo{journal}{Nature}
  \textbf{\bibinfo{volume}{427}}, \bibinfo{pages}{426} (\bibinfo{year}{2004}).

\bibitem[{\citenamefont{Fan et~al.}(2003)\citenamefont{Fan, Buczko, Puretzky,
  Geohegan, Howe, Pantelides, and Pennycook}}]{Fan03}
\bibinfo{author}{\bibfnamefont{X.}~\bibnamefont{Fan}} {\it et al.},
  \bibinfo{journal}{Phys. Rev. Lett.}
  \textbf{\bibinfo{volume}{90}}, \bibinfo{pages}{145501}
  (\bibinfo{year}{2003}).

\bibitem[{\citenamefont{Zhang et~al.}(2004)\citenamefont{Zhang, Wells, Gong,
  and Zhang}}]{Zhang04}
\bibinfo{author}{\bibfnamefont{Q.-M.} \bibnamefont{Zhang}}  {\it et al.},
  \bibinfo{journal}{Phys. Rev. B} \textbf{\bibinfo{volume}{69}},
  \bibinfo{pages}{205413} (\bibinfo{year}{2004}).
  
\bibitem[{\citenamefont{Bengaard et~al.}(2002)\citenamefont{Bengaard,
  N{\o}rskov, Sehested, Clausen, Nielsen, Molenbroek, and
  Rostrup-Nielsen}}]{Bengaard02}
\bibinfo{author}{\bibfnamefont{H.~S.} \bibnamefont{Bengaard}} {\it et al.},
  \bibinfo{journal}{J. Catal.}
  \textbf{\bibinfo{volume}{209}}, \bibinfo{pages}{365} (\bibinfo{year}{2002}).

\bibitem[{\citenamefont{Abild-Pedersen
  et~al.}(2006)\citenamefont{Abild-Pedersen, N{\o}rskov, Rostrup-Nielsen,
  Sehested, and Helveg}}]{Abild-Pedersen06}
\bibinfo{author}{\bibfnamefont{F.}~\bibnamefont{Abild-Pedersen}} {\it et al.},
  \bibinfo{journal}{Phys. Rev. B} \textbf{\bibinfo{volume}{73}},
  \bibinfo{pages}{115419} (\bibinfo{year}{2006}).  





\bibitem[{\citenamefont{Vitos et~al.}(1998)\citenamefont{Vitos, Ruban, Skriver,
  and Koll\'ar}}]{Vitos98}
\bibinfo{author}{\bibfnamefont{L.}~\bibnamefont{Vitos}}  {\it et al.},
  \bibinfo{journal}{Surf. Sci.} \textbf{\bibinfo{volume}{411}},
  \bibinfo{pages}{186} (\bibinfo{year}{1998}).

\bibitem[{\citenamefont{Baletto and Ferrando}(2005)}]{Baletto05}
\bibinfo{author}{\bibfnamefont{F.}~\bibnamefont{Baletto}} \bibnamefont{and}
  \bibinfo{author}{\bibfnamefont{R.}~\bibnamefont{Ferrando}},
  \bibinfo{journal}{Rev. Mod. Phys.} \textbf{\bibinfo{volume}{77}},
  \bibinfo{pages}{371} (\bibinfo{year}{2005}).

\bibitem[{\citenamefont{Baroni~{\it et al.}}()}]{QE}
\bibinfo{author}{\bibfnamefont{S.}~\bibnamefont{Baroni~{\it et al.}}},
  \bibinfo{note}{http://www.quantum-espresso.org/}.

\bibitem[{\citenamefont{Perdew et~al.}(1996)\citenamefont{Perdew, Burke, and
  Ernzerhof}}]{Perdew96}
\bibinfo{author}{\bibfnamefont{J.~P.} \bibnamefont{Perdew}},
  \bibinfo{author}{\bibfnamefont{K.}~\bibnamefont{Burke}}, \bibnamefont{and}
  \bibinfo{author}{\bibfnamefont{M.}~\bibnamefont{Ernzerhof}},
  \bibinfo{journal}{Phys. Rev. Lett.} \textbf{\bibinfo{volume}{77}},
  \bibinfo{pages}{3865} (\bibinfo{year}{1996}).

\bibitem[{\citenamefont{Vanderbilt}(1990)}]{Vanderbilt90}
\bibinfo{author}{\bibfnamefont{D.}~\bibnamefont{Vanderbilt}},
  \bibinfo{journal}{Phys. Rev. B} \textbf{\bibinfo{volume}{41}},
  \bibinfo{pages}{R7892} (\bibinfo{year}{1990}).

\bibitem[{\citenamefont{Pasquarello et~al.}(1992)\citenamefont{Pasquarello,
  Laasonen, Car, Lee, and Vanderbilt}}]{pasquarello92}
\bibinfo{author}{\bibfnamefont{A.}~\bibnamefont{Pasquarello}}  {\it et al.},
  \bibinfo{journal}{Phys. Rev. Lett.} \textbf{\bibinfo{volume}{69}},
  \bibinfo{pages}{1982} (\bibinfo{year}{1992});
\bibinfo{author}{\bibfnamefont{K.}~\bibnamefont{Laasonen}}  {\it et al.},
  \bibinfo{journal}{Phys. Rev. B} \textbf{\bibinfo{volume}{47}},
  \bibinfo{pages}{10142} (\bibinfo{year}{1993}).

\bibitem[{\citenamefont{Mills et~al.}(1995)\citenamefont{Mills, J\'{o}nsson,
  and Schenter}}]{Mills95}
\bibinfo{author}{\bibfnamefont{G.}~\bibnamefont{Mills}},
  \bibinfo{author}{\bibfnamefont{H.}~\bibnamefont{J\'{o}nsson}},
  \bibnamefont{and} \bibinfo{author}{\bibfnamefont{G.~K.}
  \bibnamefont{Schenter}}, \bibinfo{journal}{Surf. Sci.}
  \textbf{\bibinfo{volume}{324}}, \bibinfo{pages}{305} (\bibinfo{year}{1995});
\bibinfo{author}{\bibfnamefont{G.}~\bibnamefont{Henkelman}},
  \bibinfo{author}{\bibfnamefont{B.~P.} \bibnamefont{Uberuaga}},
  \bibnamefont{and}
  \bibinfo{author}{\bibfnamefont{H.}~\bibnamefont{J\'{o}nsson}},
  \bibinfo{journal}{J. Chem. Phys.} \textbf{\bibinfo{volume}{113}},
  \bibinfo{pages}{9901} (\bibinfo{year}{2000}).

\bibitem[{\citenamefont{Abild-Pedersen
  et~al.}(2007)\citenamefont{Abild-Pedersen, Greeley, Studt, Rossmeisl, Munter,
  Moses, Skúlason, Bligaard, and N{\o}rskov}}]{Abild-Pedersen07}
\bibinfo{author}{\bibfnamefont{F.}~\bibnamefont{Abild-Pedersen}} {\it et al.},
  \bibinfo{journal}{Phys. Rev. Lett.} \textbf{\bibinfo{volume}{99}},
  \bibinfo{pages}{016105} (\bibinfo{year}{2007}).

\bibitem[{\citenamefont{Todorova et~al.}(2002)\citenamefont{Todorova, Li,
  Ganduglia-Pirovano, Stampfl, Reuter, and Scheffler}}]{Todorova02}
\bibinfo{author}{\bibfnamefont{M.}~\bibnamefont{Todorova}} {\it et al.},
  \bibinfo{journal}{Phys. Rev. Lett.} \textbf{\bibinfo{volume}{89}},
  \bibinfo{pages}{096103} (\bibinfo{year}{2002}).

\bibitem[{\citenamefont{Rodriguez-Manzo
  et~al.}(2007)\citenamefont{Rodriguez-Manzo, Terrones, Terrones, Kroto, Sun,
  and Banhart}}]{Rodriguez-Manzo07}
\bibinfo{author}{\bibfnamefont{J.~A.} \bibnamefont{Rodriguez-Manzo}} {\it et al.},
  \bibinfo{journal}{Nat. Nanotechnol.} \textbf{\bibinfo{volume}{2}},
  \bibinfo{pages}{307} (\bibinfo{year}{2007}).

\end{thebibliography}
\end{document}